\documentclass[twoside,slac_one]{revtex4}
\usepackage{graphicx}
\usepackage{fancyhdr}
\usepackage{amsmath} 
\usepackage{bm}
\usepackage{amsxtra}
\usepackage{amssymb}
\usepackage{amsthm}
\usepackage{latexsym}
\usepackage{lscape}
\usepackage{xspace}    

\newcommand{\bb}{\ensuremath{b \overline{b}}\xspace}

\begin{document}

\title{Measurement of inclusive $b$-quark production at $\sqrt{s}=$7~TeV with the CMS experiment}

%

\author{G. Tinti}
\affiliation{Department of Physics \& Astronomy, University of Kansas, Lawrence, KS, USA}

\begin{abstract}
Measurements performed by the CMS experiment of the cross section for inclusive $b$-quark production in proton-proton collisions at $\sqrt{s}= 7$~TeV are presented. The measurements are based on different methods, such as inclusive jet measurements with secondary vertex tagging or selecting a sample of events containing jets and at least one muon, where the transverse momentum of the muon with respect to the closest jet axis discriminates $b$-events from the background. The results are compared with predictions based on perturbative QCD calculations at leading and next-to-leading order.
\end{abstract}

\maketitle

\thispagestyle{fancy}

\section{Introduction}
Cross section measurements of \bb production at $\sqrt{s}=7~\textrm{TeV}$ are interesting as a proof of perturbative Quantum Chromodynamics (QCD) at both leading and next-to-leading order (NLO). At the Large Hadron Collider (LHC) energies, all the three \bb production mechanisms (flavor creation, gluon splitting and flavor excitation) are achievable. 

The Compact Muon Solenoid (CMS) experiment is a general purpose experiment~\cite{tdr}. The main detector components used in the analyses presented here are the silicon tracker and the muon system. The silicon tracker has pseudorapidity coverage up to $|\eta|<2.5$ and provides an impact parameter resolution of $\approx 15~\mu$m. Due to the 3.8~T magnetic field, the transverse momentum ($p_T$) resolution is $\approx 1.5\%$ for particles with transverse momentum below 100~GeV. The muon system has a coverage up to $|\eta|<2.4$. Information from the calorimeters is used to fully reconstruct jets.   

The measurements presented in this paper use early CMS data (2010) as the trigger paths used in the following analyses are quite loose (low transverse muon or jet momentum). With the increase of the instantaneous luminosity and the validation of the Standard Model predictions at the LHC scale, tighter triggers have started to be applied and $b$-quark production measurements with newer data are not available at low momenta. 

The paper presents three cross section analyses: \bb cross section using dimuons is presented in Section~\ref{dimuonsec}, the open beauty production using single muons is presented in Section~\ref{ptrelsec} and the $b$-jet inclusive cross section is presented in Section~\ref{bjetsec}. Section~\ref{bbarsec} describes a measurement of the angular correlations between the $B\overline{B}$ hadrons. This analysis is sensitive to the individual production mechanisms. Conclusions are presented in Section~\ref{conclusions}.   

\section{Correlated $b \bar{b}$ production with dimuons}\label{dimuonsec}

The cross section for \bb production with both $b$-quarks decaying into muons has been recently measured~\cite{dimuonpaper} by CMS using $\mathcal{L}=28~\textrm{pb}^{-1}$ of data. The flavor composition of the dimuon sample in data has been determined by using the muon transverse impact parameter with respect to primary vertex ($d_{xy}$) as a discriminating variable. Both muons need to have the transverse momentum $p^\mu_T>4$~GeV, pseudorapidity $|\eta^\mu|<2.1$ and to be associated to the same primary vertex. Cuts on invariant mass of the dimuon pair, $M_{\mu \mu}$, remove backgrounds from the $Z$ ($M_{\mu \mu}>70$~GeV), the $\Upsilon$ ($8.9<M_{\mu \mu}<10.6$~GeV) and charmonium resonances and sequential decays ($M_{\mu \mu}<5$~GeV).

Single muon events in the simulation are classified as: 
\begin{itemize}
\item \textbf{B}: if the muon comes from a $b$-quark (either directly or with sequential decays)
\item \textbf{C}: if the muon comes from a $c$-quark 
\item \textbf{P}: muons from prompt tracks or fakes 
\item \textbf{D}: meson decay in flight (muon from light quarks).
\end{itemize}
The impact parameter distributions for the \textbf{B}, \textbf{C} and \textbf{D} categories have been computed from simulation. The distribution for the \textbf{P} category is extracted from data by looking at $\Upsilon$ decays.

\begin{figure}[h]
\centering
\includegraphics[width=80mm]{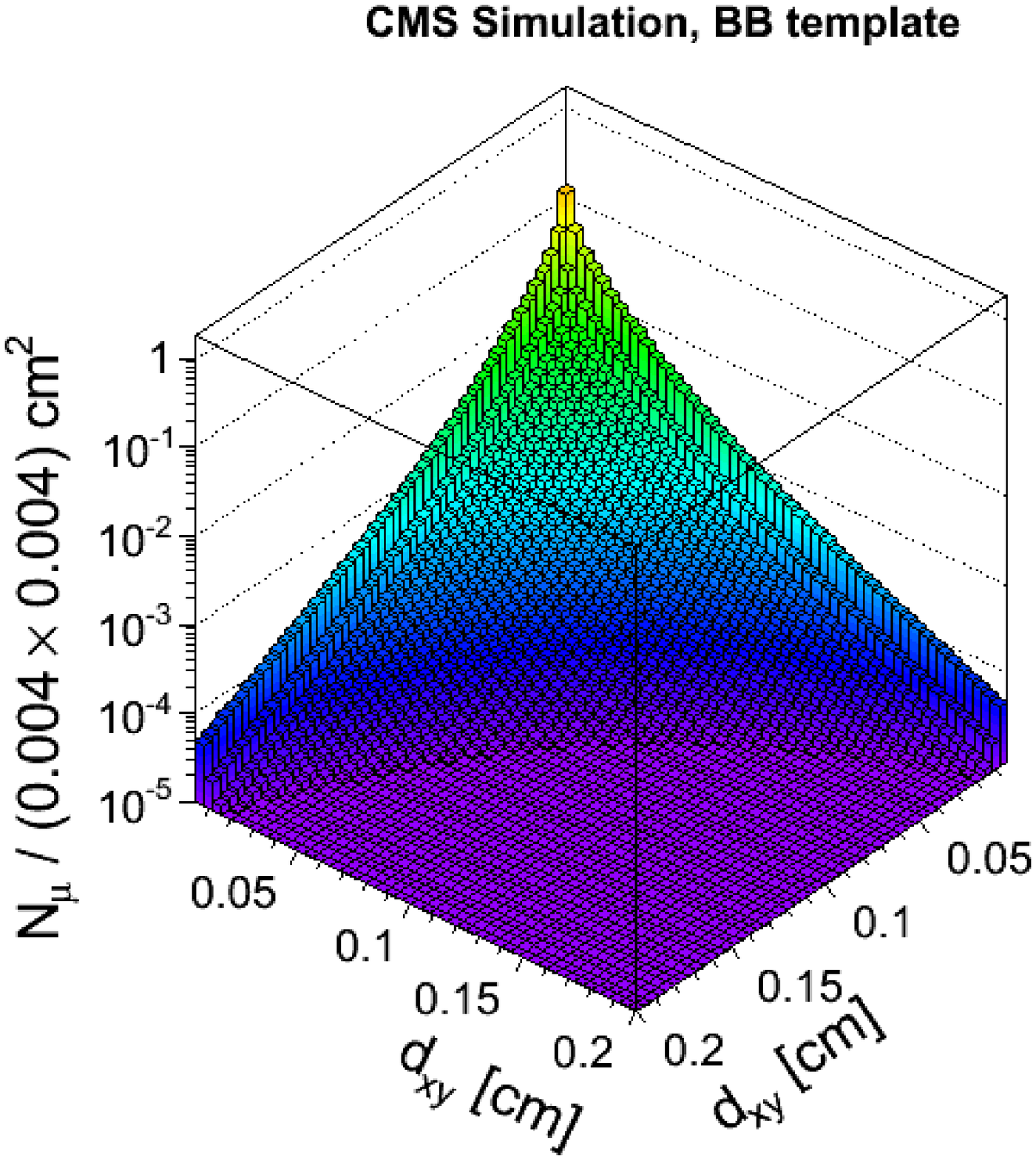}
\includegraphics[width=80mm]{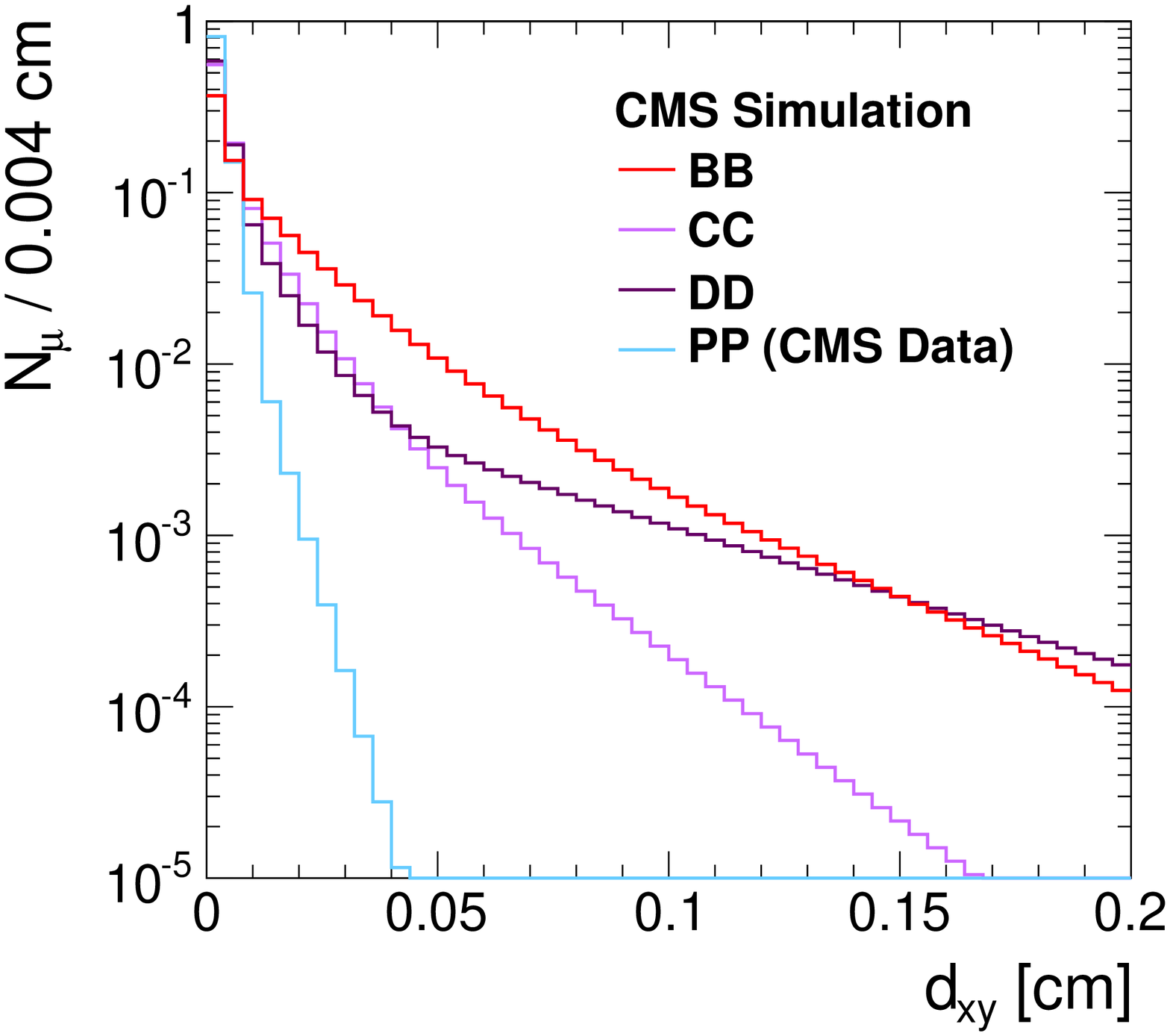}
\caption{Left: example of a 2D template (\textbf{BB}) as constructed from symmetrization of the 1D distribution for the muon impact parameter. Right: 1D projection of the 2D templates for the categories where both muons come from the same source. The difference in shape between the \textbf{BB} template and the others can be seen.}
\label{2D}
\end{figure}

2D dimuon templates for the muon impact parameter (\textbf{BB}, \textbf{CC}, \textbf{PP}, \textbf{DD}) and combinations (\textbf{BD}, \textbf{BC}, \textbf{CD}...) are built from the 1D distributions of the impact parameter for each muon. The 2D symmetrized distributions are populated in the $T_{12, ij}$ bin according to the weight:
\begin{equation}
\nonumber
T_{12, ij}=\frac{S_{1,i} \cdot S_{2,j} + S_{1,j} \cdot S_{2,i}}{2},
\end{equation}
which takes into account the bin content $S_{1(2),i}$ and $S_{1(2),j}$ of the 1D distribution for the muon $1(2)$.
2D templates for data are filled by choosing the first and second muon in random order. 
Figure~\ref{2D} (left) shows one 2D template obtained in the case both muons come from $b$-quarks (\textbf{BB} category). Figure~\ref{2D} (right) shows the 1D projection of the 2D templates. Only processes in which both muons come from the same sources are shown in the plot.  

\begin{figure}[h]
\centering
\includegraphics[width=80mm]{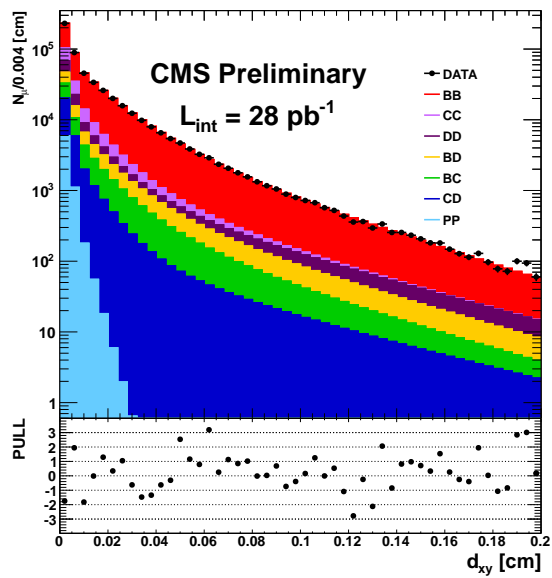}
\caption{1D projection of the fit results for all templates considered superimposed with the data. The pull distribution of the data and the simulation is shown in the plot at the bottom.}
\label{2Dresults}
\end{figure}

A binned maximum likelihood fit using the 2D distributions is performed to the data in order to extract the \textbf{BB} fraction content. Templates for both muons from the same source or each muon from different sources are used. The templates for the \textbf{BP}, \textbf{CP} and \textbf{PD} have been neglected as they cannot be resolved in the fit and have a small contribution. Constraints are applied to the binned likelihood fit to maintain the ratios \textbf{BC/BB}, \textbf{BD/BB} and \textbf{CD/CC} at the simulation predictions. Figure~\ref{2Dresults} shows the 1D projections of the 2D distributions for data and for all the templates after the fit. The fraction of \textbf{BB} obtained by the fit is 65\%. 

The integrated production cross section in the kinematic region ($p_T^\mu > 4~\textrm{GeV},|\eta^\mu|<2.1$) has been evaluated:
\begin{equation}
\nonumber
\sigma ( pp \to b \overline{b} X \to \mu \mu Y ) = 26.18 \pm 0.14\:(\textrm{stat.}) \pm 2.82 \;(\textrm{syst.}) \pm 1.05\; (\textrm{lumi.})~\textrm{nb}.
\end{equation}
The largest systematic uncertainty in the analysis is from trigger efficiency ($\approx 8.3\%$).
The result can be compared with the PYTHIA 6.4~\cite{PYTHIA} prediction:
\begin{equation}
\nonumber
\sigma_{\textrm{PYTHIA}}= 48.2~\textrm{nb}
\end{equation}
and the MC@NLO~\cite{MC@NLO} one:
\begin{equation}
\nonumber
\sigma_{\textrm{MC@NLO}} = 19.95 \pm 0.46\; (\textrm{stat.}) \phantom{1}^{+4.68}_{-4.33} \;(\textrm{scale+pdf+m$_{b}$})~\textrm{nb},
\end{equation}
where pdf stands for the parton distribution function.
The central value of the measurement is between the central values for the PYTHIA and MC@NLO predictions. Data and MC@NLO agree within uncertainties. 

\section{Open beauty production with muons} \label{ptrelsec}
An early data analysis, performed with data corresponding to an integrated luminosity of $\mathcal{L}=85~\textrm{nb}^{-1}$, measures the inclusive beauty production using single muons~\cite{ptrelpaper}. Signal events are discriminated using the muon transverse momentum relative to the jet direction $p^{\textrm{rel}}_T$, defined as:
\begin{equation}
\nonumber
p^{\textrm{rel}}_T=\frac{| \vec{p}_\mu \times \vec{p}_{\textrm{jet}}|}{|\vec{p}_{\textrm{jet}}|}.
\end{equation}
The $p^{\textrm{rel}}_T$ distribution is harder in $b$-events than in background events, due to the larger mass of the $b$-quark. The background events are categorized as events in which the muon come from a $c$-quark or from a light quark ($udsg$). The analysis requires the muon transverse momentum $p^\mu_T > 6$~GeV, pseudorapidity $|\eta^\mu|<2.1$, transverse impact parameter $|d_{xy}|<2$~mm and longitudinal impact parameter $|d_z|<1$~cm. Track Jets~\cite{trackjet} are used to determine jet direction. In this measurement, the full jet reconstruction is not critical, provided that the jet direction is well defined. Tracks with $p_T>300$~MeV are clustered in a jet using an anti-$k_T$ algorithm~\cite{antikt}, with a cone radius of $R=0.5$.  

\begin{figure}[h]
\centering
\includegraphics[width=80mm,angle=90]{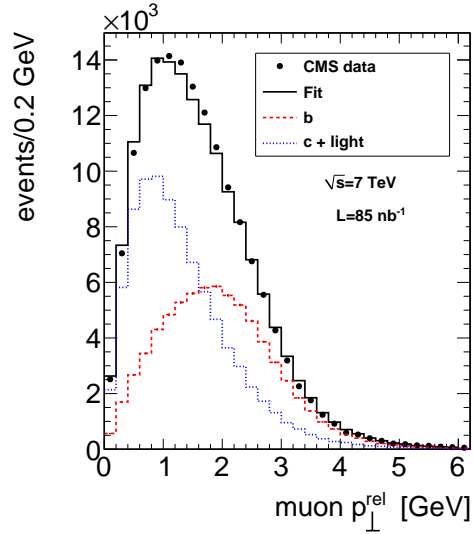}
\caption{Distribution of the $p_T^{\textrm{rel}}$ distribution for data (dots), and the fitted MC (black histogram). The separated distributions for muons from $b$-quarks and from the $c+$light quarks are also shown.}
\label{ptrel}
\end{figure}

 A binned maximum likelihood fit is performed with the $p^{\textrm{rel}}_T$ distribution based on templates. The shape of the light quark/gluon component is derived from minimum bias data by evaluating the misidentification probability for hadrons to be selected as muons. The templates for $c$ and light quark/gluon components are combined in the likelihood computation. As an overall result, the $b$-fraction returned by the fit is 46\%. Figure~\ref{ptrel} shows the distrubution of the $p_T^{\textrm{rel}}$ for data, and the distributions for the signal ($b$) and the background ($c$+light) after the fit.

\begin{figure}[h]
\centering
\includegraphics[width=70mm,angle=90]{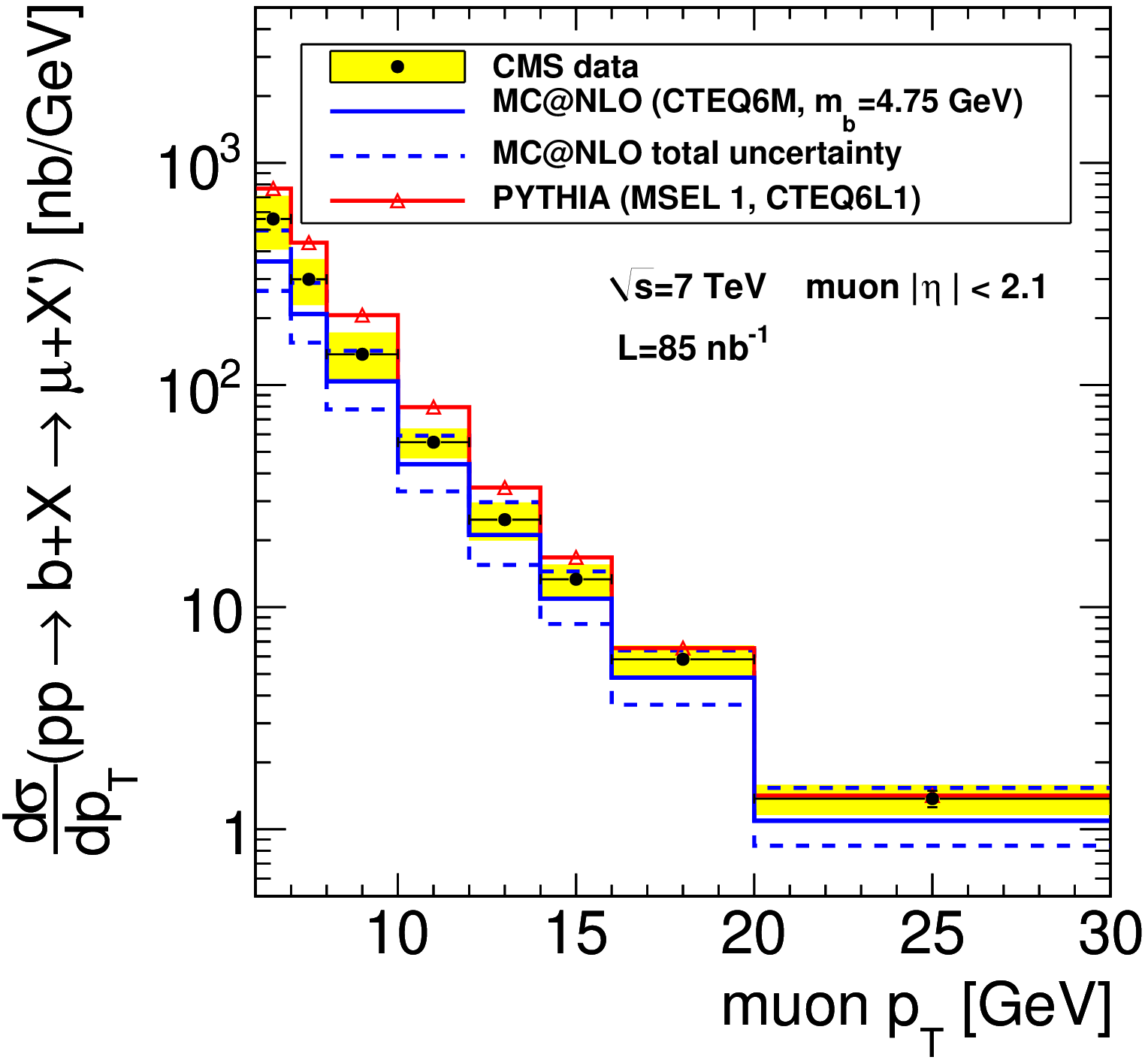}
\includegraphics[width=70mm,angle=90]{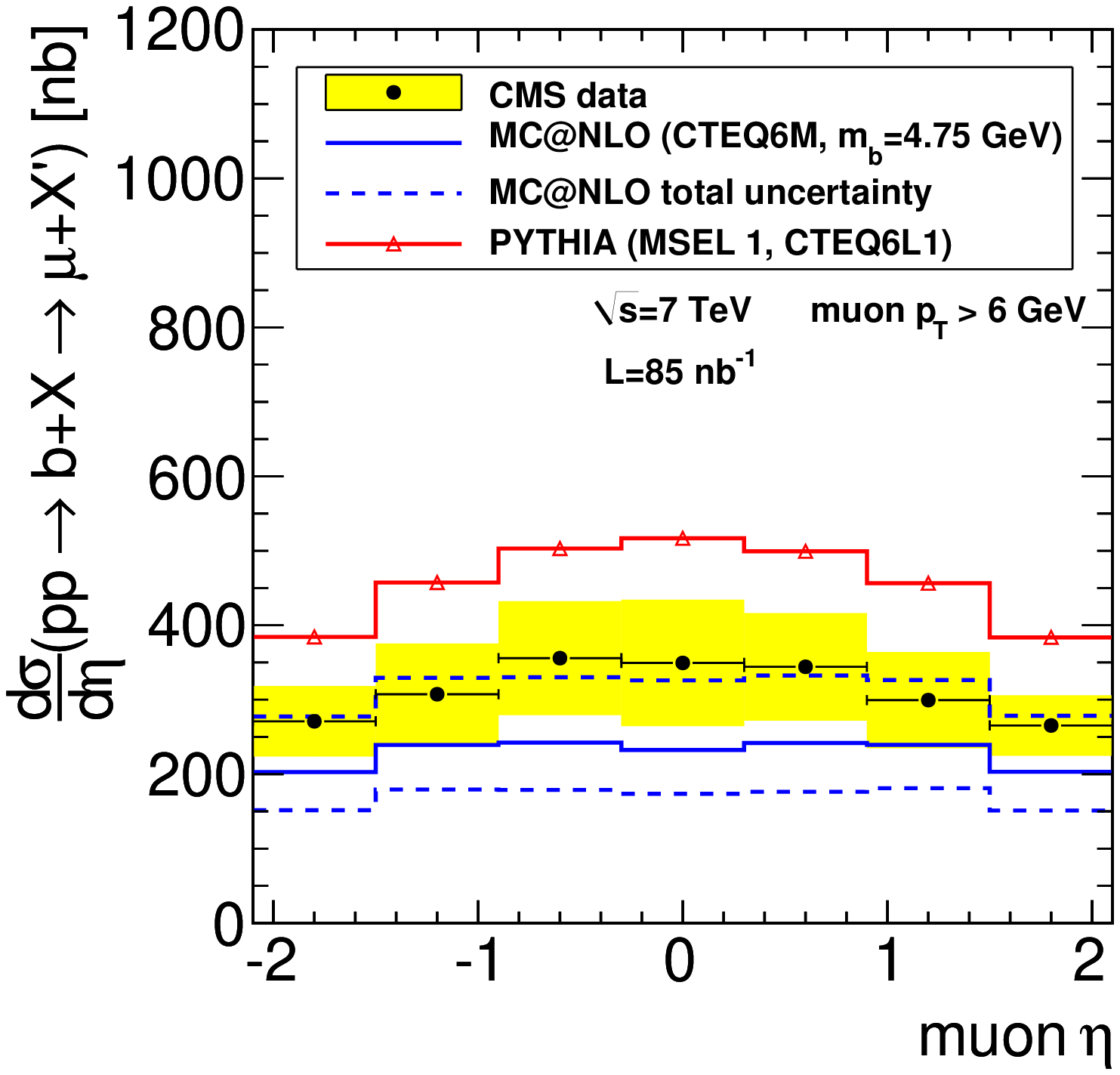}
\caption{Differential cross section as a function of the muon $p_T$ (left) and of the muon pseudorapidity (right). The black points are the CMS data. The uncertainty on the measurement is shaded in yellow. The 11\% luminosity uncertanty is not included. The MC@NLO (with uncertainties) and PYTHIA predictions are shown.}
\label{ptrelresults}
\end{figure}

The analysis has been performed integrating over the muon kinematic phase space to obtain the integrated cross section and in bins of muon $p_T^\mu$ and $\eta^\mu$ to obtain the differential production cross-sections. 
The measured integrated cross section in the kinematic range $p_T^\mu >6$~GeV and $|\eta^\mu |<2.1$ is:
\begin{equation}
\nonumber
\sigma = 1.32 \pm 0.01 \; \textrm{(stat.)} \pm 0.30 \; \textrm{(syst.)} \pm 0.15 \;\textrm{(lumi.)}~\mu \textrm{b}.
\end{equation}
The major systematic uncertainties of the measurement come from the $b$-quark's $p^{\textrm{rel}}_T$ shape uncertainty (21\%) and the luminosity uncertainty (11\%). The prediction from PYTHIA 6.4 for the same process is:
\begin{equation}
\nonumber
\sigma_{\textrm{PYTHIA}}=1.9~\mu \textrm{b},
\end{equation}
while the prediction for MC@NLO 3.4 is:
\begin{equation}
\nonumber
\sigma_{\textrm{MC@NLO}} = 0.95^{+0.41}_{-0.21} \; (\textrm{scale}) \pm 0.09 \; (m_b) \pm 0.05 \; (\textrm{pdf})~\mu \textrm{b}.
\end{equation}
The measured cross section has a value between the two predictions and is compatible with MC@NLO within uncertainties. Figure~\ref{ptrelresults} shows the differential cross section as a function of the muon $p_T^\mu$ (left) and the muon $\eta^\mu$ (right). Data are compared to the MC@NLO and the PYTHIA predictions. As in the cases seen before, data and MC@NLO agree within uncertainties.

\section{Inclusive $b$-jet production}\label{bjetsec}

\begin{figure}[h]
\centering
\includegraphics[width=80mm]{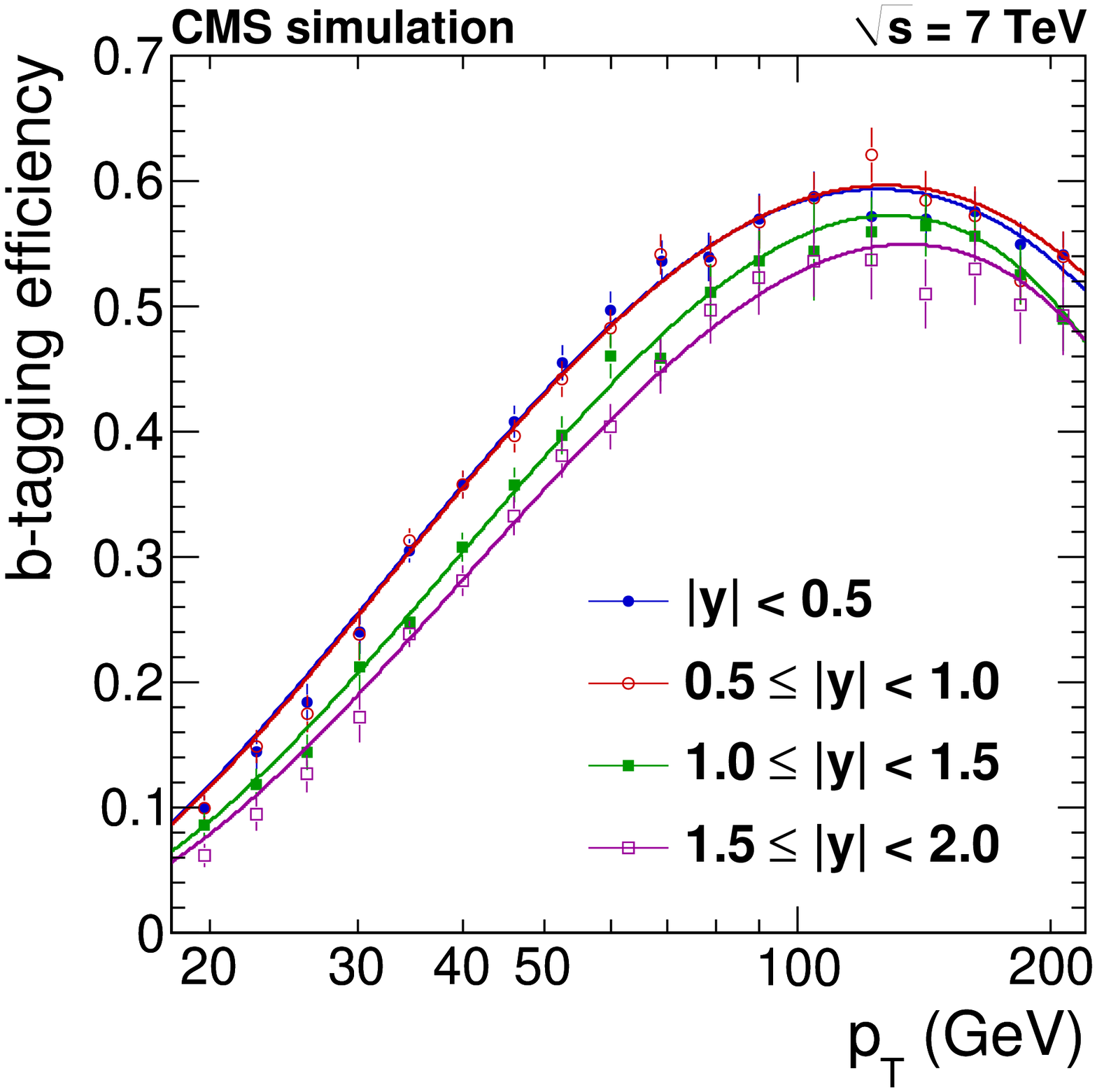}
\includegraphics[width=80mm]{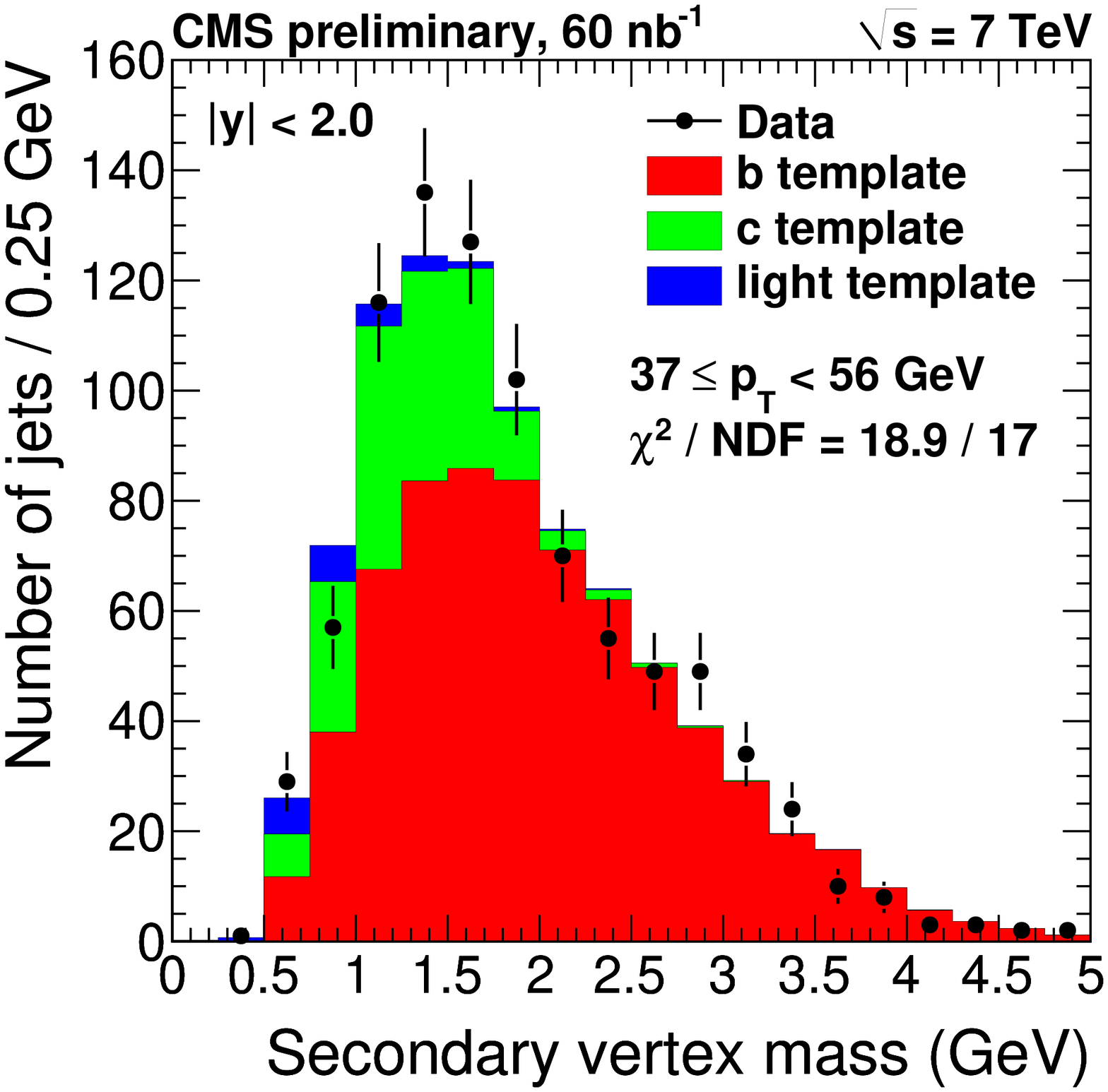}
\caption{Left: $b$-tagging efficiency from the secondary vertex tagger as a function of the jet transverse momentum in different jet rapidity bins. The plot is from simulation. Right: Example of a secondary vertex mass fit. The fit is performed in order to obtain the purity of the $b$-tagged jets from data.}
\label{btageffpur}
\end{figure}

Another early data analysis~\cite{bjetpaper} measures the inclusive $b$-jet production using $\mathcal{L}=60~\textrm{nb}^{-1}$ of data.
 In this analysis the jet reconstruction is important and the full detector information is used to reconstruct Particle Flow jets~\cite{pfjets}. Jet reconstruction is performed by using an anti-$k_T$ algorithm, with $R=0.5$. The jets are required to have: $18< p^{\textrm{jet}}_T < 300$~GeV and rapidity $|y|<2$. $b$-jets are identified using a secondary vertex tagger~\cite{btag}: the secondary vertex finder is seeded by at least 3 charged particle tracks and the 3D decay length significance cut is applied. The $b$-tagging efficiency as a function of the transverse momentum $p^{\textrm{jet}}_T$ and rapidity $y^{\textrm{jet}}$ is evaluated from simulation. Data-driven corrections use the $p^{\textrm{rel}}_T$ distributions to determine the data/simulation scaling factors as a function of the jet $p_T$ and $y$ to apply to the $b$-tagging efficiency from the secondary vertex tagger. The purity of the $b$-jet selection is estimated from data by fitting the secondary vertex mass distribution after the selection. Figure~\ref{btageffpur} shows the $b$-tagging efficiency from simulation (left) for different jet rapidity regions and the distribution of the secondary vertex mass for data and simulation (right).

\begin{figure}[h]
\centering
\includegraphics[width=80mm]{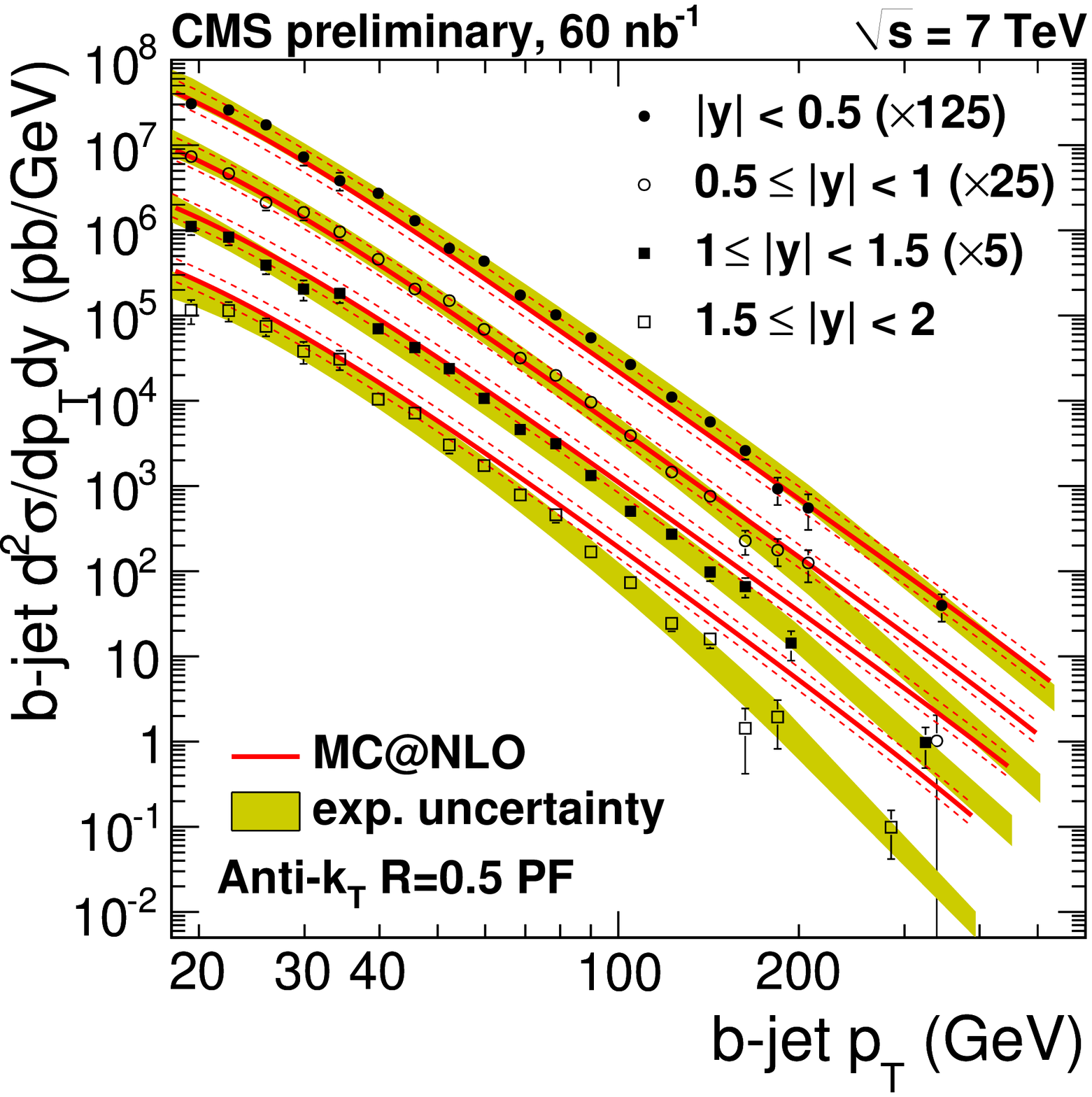}
\includegraphics[width=80mm]{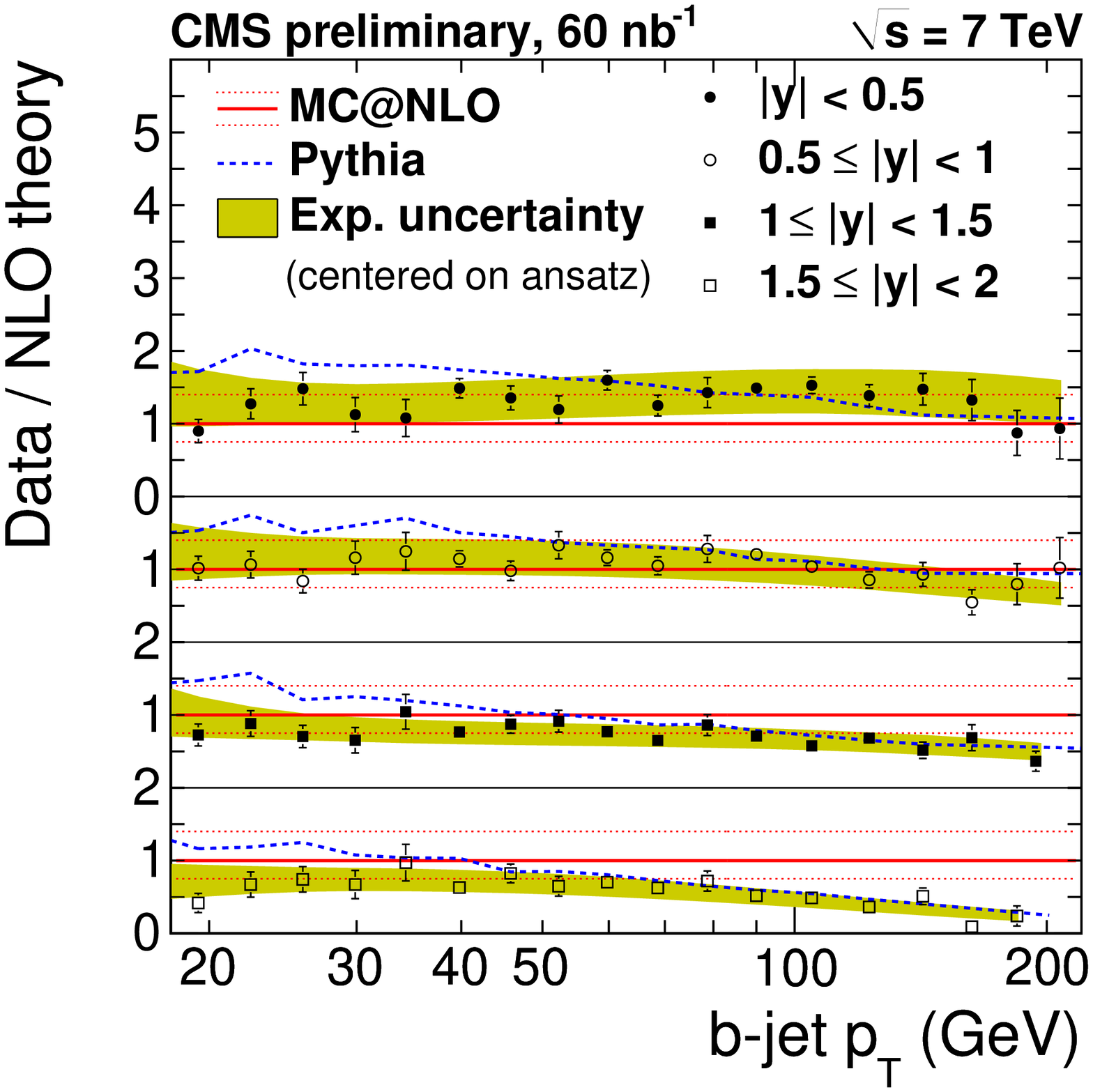}
\caption{Left: Measured differential $b$-jet cross section as a function of the $b$-jet $p_T$ in different $b$-jet rapidity bins. The MC@NLO prediction is also shown. Right: Data/simulation prediction comparison. Both the MC@NLO and PYTHIA predictions are shown.}
\label{btagresults}
\end{figure}

Figure~\ref{btagresults} (left) shows the double differential production cross section as a function of the $b$-jet transverse momentum in bins of jet rapidity. The plot shows the cross section as measured in data, and as predicted by MC@NLO. MC@NLO describes the overall fraction of $b$-jets, but there are shape differences in $p^{\textrm{jet}}_T$ and $y^{\textrm{jet}}$. The main systematic uncertainties in the measurement come from the $b$-tag efficiency (20\%), the jet energy scale (5\%) and the luminosity (11\%). Figure~\ref{btagresults} (right) shows the ratio of the measured cross section to the predicted MC@NLO one as a function of jet transverse momentum. Data and MC@NLO agree within uncertainties. The PYTHIA prediction divided by the MC@NLO one is also shown. For $b$-jet transverse momentum greater than 30~GeV, the PYTHIA prediction agrees fairly well with the data. 

\section{$B\overline{B}$ angular correlations}\label{bbarsec}
The analysis in~\cite{bbarpaper} measures the angular correlations between the $B\overline{B}$ hadrons, using $\mathcal{L}=3.1$~pb$^{-1}$ of data. The angular correlations between $B\overline{B}$ are sensitive to the production mechanism, as the pairs produced by the gluon splitting are expected to have a small separation, while the ones produced from flavor excitations are produced at large separation. 

\begin{figure}[h]
\centering
\includegraphics[width=80mm]{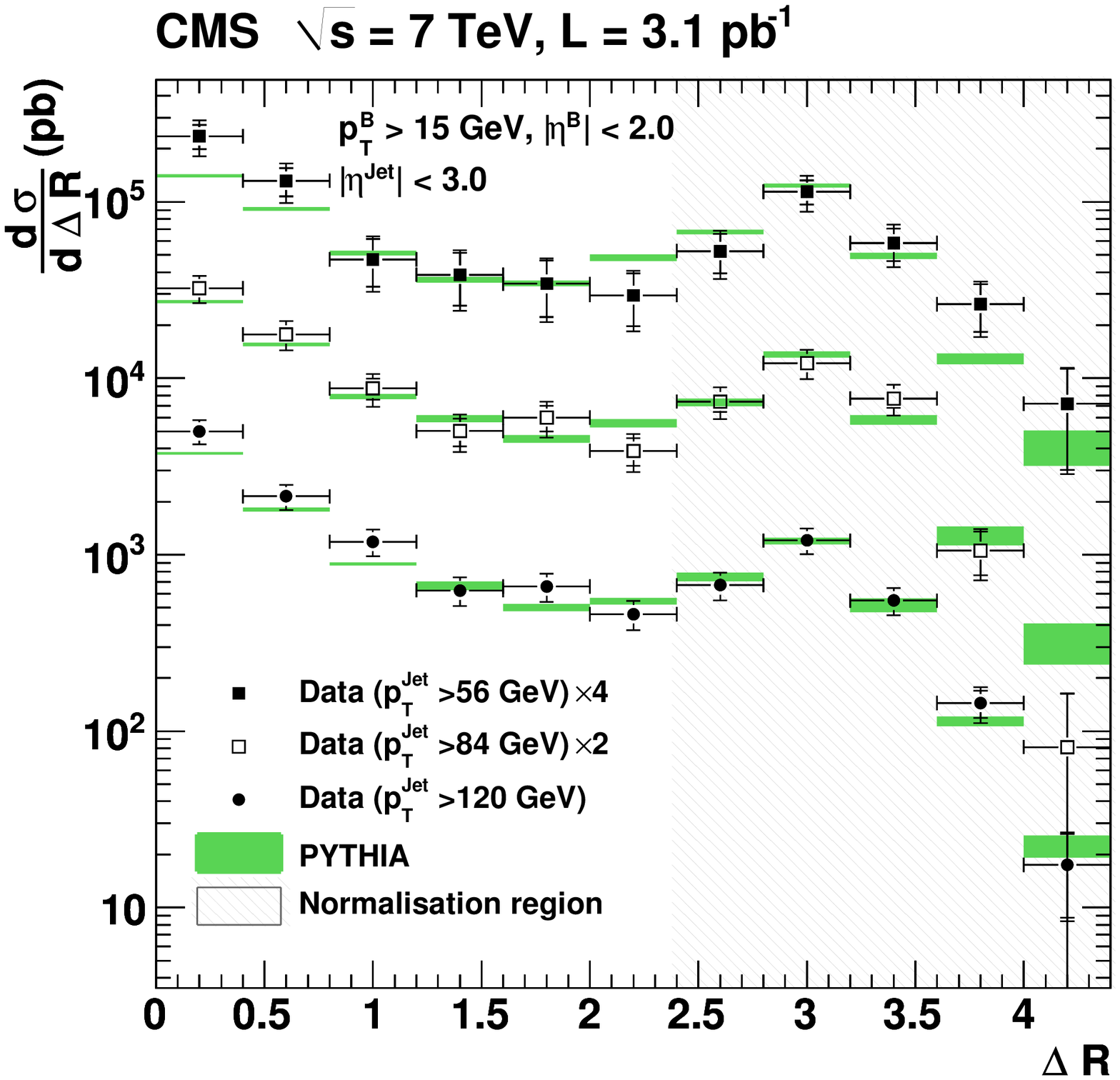}
\includegraphics[width=80mm]{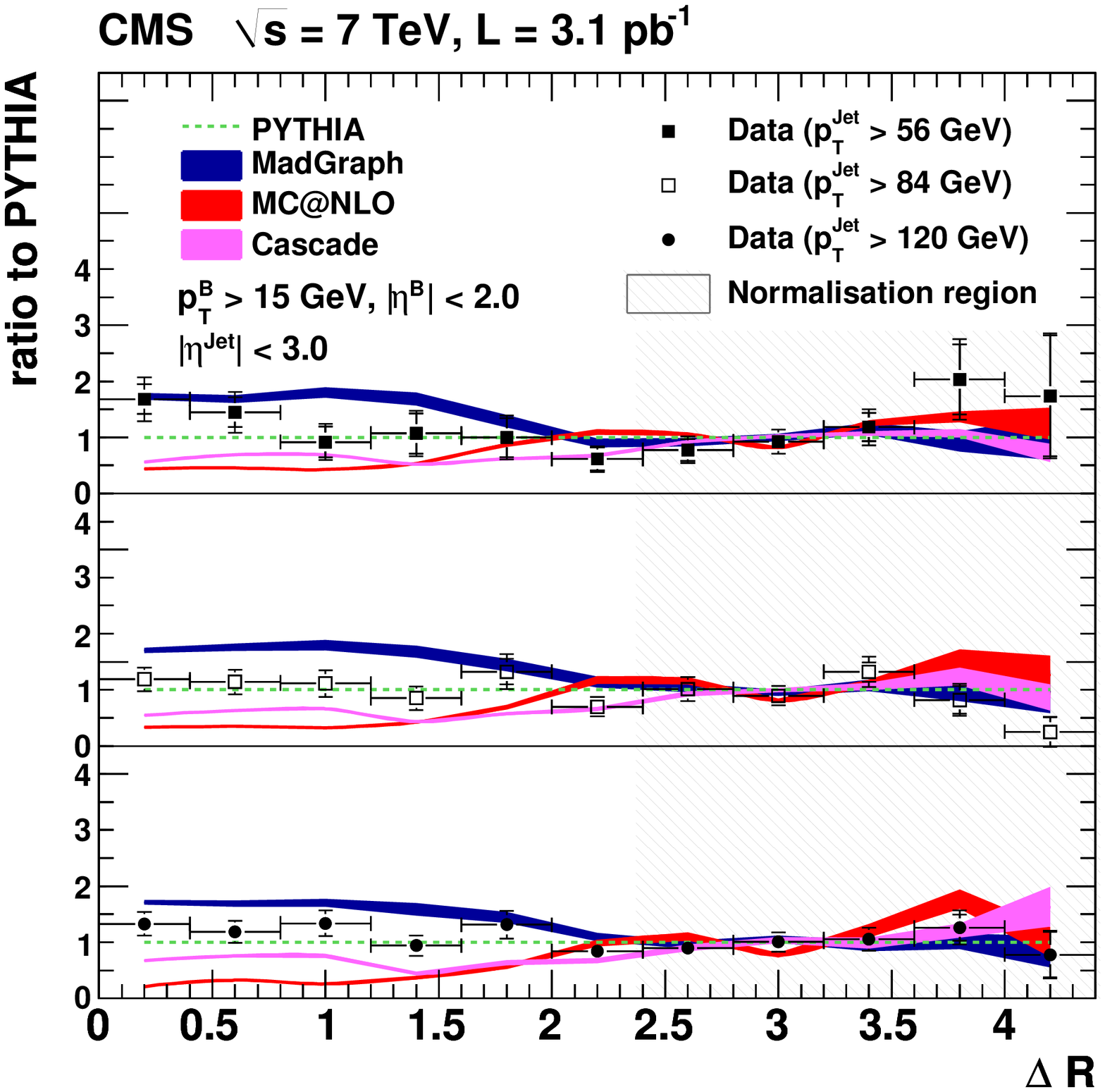}
\caption{Left: differential $B\overline{B}$ production cross section as a function of $\Delta R$. The measurements are presented in 3 bins of leading jet $p_T$. The data measurements are presented, as well as the PYTHIA prediction. The normalization region is shown. Right: ratio of the data differential $B\overline{B}$ cross section to the PYTHIA prediction as a function of $\Delta R$. The MC@NLO prediction is also compared to PYTHIA.}  
\label{bbbarresults}
\end{figure}

The analysis measures $\Delta R=\sqrt{(\Delta \eta)^2 + (\Delta \phi)^2}$ between the flight directions of the two $B$-hadrons. A single jet trigger is used; the jet is only used to set the energy scale and does not enter the $B$-hadron reconstruction. The secondary vertices are reconstructed using an inclusive vertex finder. Secondary vertices can be reconstructed even if the two $B$-hadrons are in the same jet. The differential production cross section in bins of the opening cone $\Delta R$ of the $B\overline{B}$ pairs has been measured, as shown in Figure~\ref{bbbarresults} (left). The data are compared to the PYTHIA 6.4 predictions. The differential cross section at large opening angles, which is less sensitive to theoretical uncertainties, is used for normalization. The measurement shows that the contribution at small opening angles is even larger than the contribution at large opening angles. The PYTHIA prediction follows the shape of the distribution, but underestimates the contribution at lower angles (where the contribution of the gluon spitting is predominant). From Figure~\ref{bbbarresults} (right), it is possible to see the MC@NLO comparison (presented as a ratio to the PYTHIA prediction).
MC@NLO does not describe the data well, particularly at small $\Delta R$, where it underestimates the contribution.

\section{Conclusions}\label{conclusions}
The excellent performances of the LHC accelerator and the CMS experiment have allowed for a fast confirmation of the Standard Model predictions even with early data. The paper presented here shows three $b$-quark production cross section measurements: the \bb production using dimuons, open beauty production using single muons and the inclusive $b$-jet production. Inclusive measurements are in overall agreement with perturbative QCD and in particular with the MC@NLO calculations. In the measurement of the production cross section of $B\overline{B}$ as a function of the angular separation, MC@NLO does not describe data well at small $B \overline{B}$ opening angles (where the gluon splitting contribution is predominant).

\begin{acknowledgments}
We wish to congratulate our colleagues in the CERN accelerator departments for the excellent performance of the LHC machine. We thank the technical and administrative staff at CERN and other CMS institutes, and acknowledge support from: FMSR (Austria); FNRS and FWO (Belgium); CNPq, CAPES, FAPERJ, and FAPESP (Brazil); MES (Bulgaria); CERN; CAS, MoST, and NSFC (China); COLCIENCIAS (Colombia); MSES (Croatia); RPF (Cyprus); Academy of Sciences and NICPB (Estonia); Academy of Finland, MEC, and HIP (Finland); CEA and CNRS/IN2P3 (France); BMBF, DFG, and HGF (Germany); GSRT (Greece); OTKA and NKTH (Hungary); DAE and DST (India); IPM (Iran); SFI (Ireland); INFN (Italy); NRF and WCU (Korea); LAS (Lithuania); CINVESTAV, CONACYT, SEP, and UASLP-FAI (Mexico); MSI (New Zealand); PAEC (Pakistan); SCSR (Poland); FCT (Portugal); JINR (Armenia, Belarus, Georgia, Ukraine, Uzbekistan); MST, MAE and RFBR (Russia); MSTD (Serbia); MICINN and CPAN (Spain); Swiss Funding Agencies (Switzerland); NSC (Taipei); TUBITAK and TAEK (Turkey); STFC (United Kingdom); DOE and NSF (USA). 
\end{acknowledgments}

\end{document}